\documentclass[trackchanges,twocolumn]{aastex701}

\usepackage{bm,enumitem}
\usepackage{newtxmath,newtxtext}

\begin{document}
\title{Positive AGN Feedback Enhances Star Formation in Starburst Dwarf Galaxies}

\author[0009-0003-1442-105X]{Tingfang Su}
\affiliation{Astrophysics Division, Shanghai Astronomical Observatory, Chinese Academy of Sciences, 80 Nandan Road, Shanghai  200030, P.R.China.}
\affiliation{University of Chinese Academy of Sciences, 19A Yuquan Road, Beijing 100049, P.R.China.}
\email{tingfangsu@shao.ac.cn}

\author[0000-0001-9658-0588]{Suoqing Ji}
\affiliation{Center for Astronomy and Astrophysics and Department of Physics, Fudan University, Shanghai 200438, P.R.China.}
\affiliation{Key Laboratory of Nuclear Physics and Ion-Beam Application (MOE), Fudan University, Shanghai 200433, P.R.China.}
\email[show]{sqji@fudan.edu.cn}

\author[0000-0003-3564-6437]{Feng Yuan}
\affiliation{Center for Astronomy and Astrophysics and Department of Physics, Fudan University, Shanghai 200438, P.R.China.}
\email[show]{fyuan@fudan.edu.cn}

\author[0009-0004-3881-674X]{Haojie Xia}
\affiliation{Astrophysics Division, Shanghai Astronomical Observatory, Chinese Academy of Sciences, 80 Nandan Road, Shanghai  200030, P.R.China.}
\affiliation{University of Chinese Academy of Sciences, 19A Yuquan Road, Beijing 100049, P.R.China.}
\email{hjxia@shao.ac.cn}

\author[0009-0006-4662-3053]{Yuxuan Zou}
\affiliation{Astrophysics Division, Shanghai Astronomical Observatory, Chinese Academy of Sciences, 80 Nandan Road, Shanghai  200030, P.R.China.}
\affiliation{University of Chinese Academy of Sciences, 19A Yuquan Road, Beijing 100049, P.R.China.}
\email{yxzou@shao.ac.cn}

\begin{abstract}
The role of active galactic nuclei (AGN) feedback in dwarf galaxies remains poorly understood, with conventional wisdom suggesting it primarily suppresses star formation. Using high-resolution MACER3D simulations that directly resolve the Bondi radius, we demonstrate that AGN feedback can significantly enhance rather than suppress star formation in starburst dwarf galaxies. Our simulations reveal that AGN feedback increases global star formation rates by approximately 25\% when comparing our models with both AGN and supernova feedback to those with only supernova feedback. This enhancement occurs through AGN-driven outflows creating compressed gas regions where efficient cooling preserves high-density gas while quickly radiating away thermal energy, creating ideal conditions for star formation. This positive feedback mechanism operates in gas-rich starburst environments with efficient cooling and moderate AGN energy input ($\sim 10^{42}\,\mathrm{erg\,s^{-1}}$) that compresses gas without expelling it from the galaxy. Critically, it requires both AGN and supernova feedback working in concert: without SN feedback to regulate black hole activity, AGN outflows become too powerful and expel gas rather than compress it. Our results align with observations of the starburst dwarf galaxy Henize 2-10, where similar shock-compressed regions of enhanced star formation have been observed. These findings challenge conventional understanding of AGN feedback and suggest that AGN may play a previously unrecognized role in accelerating star formation during active phases of dwarf galaxy evolution.
\end{abstract}


\section{Introduction}  \label{sec:intro}

Star formation is a fundamental process of galaxy evolution, and baryonic feedback mechanisms are widely regarded as key regulators of this process \citep{Kennicutt2012}. 
In these mechanisms, active galactic nuclei (AGN) feedback has been extensively discussed. 
It is generally considered playing a significant role in massive galaxies, which is well-studied in both observations \citep[e.g.,][]{Fabian2012, Heckman2014, Lammers2023} and theoretical works \citep[e.g.,][]{Dubois2013, Yuan2018, Su2021}.
Supermassive black holes (SMBHs), which are universal in massive galaxies \citep{Magorrian1998, Kormendy2013}, interact with the interstellar medium (ISM) through outflows and radiation generated by accretion, thereby modulating star formation \citep{King2015}.
For decades, SMBH feedback has been considered as a crucial mechanism for quenching star formation in massive galaxies, thereby reproducing the observed sharp decline at the bright end of the galaxy luminosity function \citep{Silk2012}. Numerous numerical simulations have shown that AGN feedback can effectively quench or significantly suppress star formation in massive galaxies \citep{Croton2006,Dubois2013,Su2021,Guo2026,Zou2026}. 
However, observational evidence remains inconclusive: while AGN host galaxies are often found in the green valley, suggesting a role in quenching \citep{Schawinski2007,Silverman2008}, spatially resolved observations reveal both suppression and enhancement of star formation in the vicinity of AGN \citep{Lammers2023,Shin2019,Pak2023,Zhang2023,Venturi2023}. 
This possibility has also been explored in several analytical studies, which suggest that AGN-driven shocks may compress the ISM and enhance star formation \citep{Ishibashi2012, Silk2013, Zubovas2013, Ishibashi2013}.

In contrast, AGN feedback is usually ignored in dwarf galaxies. The lower luminosities of AGN powered by lower-mass black holes (BHs) makes the detection of AGN signatures, such as optical emission lines, X-ray emission, and radio jets, more challenging compared to massive galaxies \citep{Reines2013, Baldassare2020,Urquhart2022, Mezcua2019}. The reduced luminosities can affect their emission-line properties, causing them to deviate from luminous AGN and potentially fall outside the AGN-dominated region in standard Baldwin-Phillips-Terlevich (BPT) diagnostics, leading to misclassification. In addition, typical dwarf galaxies beyond the local universe are frequently too faint to be detected with current observational facilities and observational samples are biased toward star-forming systems. This further complicates the identification of AGN activity and hampers accurate measurements of the AGN fraction in dwarf galaxies \citep{Kaviraj2025}.
Meanwhile, supernova (SN) feedback has long been regarded as the dominant baryonic feedback process \citep{Dekel1986}.
However, there are still some problems that cannot be explained by SN feedback alone.
For example, the ``core-cusp'' problem, which describes the discrepancy between ${\rm \Lambda CDM}$-predicted cuspy dark matter profiles and observed cored profiles in dwarfs, has yet to be adequately resolved by SN feedback \citep{Governato2010, Pontzen2014}.
If the star formation rate is too low, the energy released by supernovae is insufficient to significantly modify the halo density profile, resulting in a dark matter distribution similar to that predicted by dark-matter-only simulations \citep{Bullock2017}.
Other processes, such as the ultraviolet background (UVB) from cosmic reionization and environmental effects (e.g., tidal and ram pressure stripping), also influence dwarf galaxy evolution. The UVB can heat the intergalactic medium and suppress gas accretion onto low-mass halos, reducing the abundance of low-mass galaxies \citep{Efstathiou1992,Okamoto2008}. Meanwhile, tidal fields can enhance disk instabilities and promote star formation in dwarf galaxies \citep{Williamson2016}.

Recent observations have increasingly detected AGN activity in dwarf galaxies \citep{Satyapal2007, Reines2013, Moran2014, Baldassare2017, Mezcua2024}.
Nevertheless, the role of AGN feedback in dwarf galaxies is even less well understood.
Some observations suggest that AGN feedback in dwarfs may suppress star formation or even quench galaxies \citep{Penny2018, Manzano-King2019}. Additionally, numerous simulations of dwarf galaxies report that AGN feedback can suppress star formation to varying degrees \citep{Koudmani2019, Koudmani2022, Sharma2023, Arjona-Galvez2024, Bi2025}.
However, observational evidence from \cite{Schutte2022} demonstrates that AGN-driven outflow can trigger star formation in a nearby starburst dwarf galaxy Henize 2-10. 
\cite{Hazenfratz2025} also finds modest star formation enhancement due to AGN feedback in isolated simulations.
The impact of AGN feedback on star formation remains uncertain in both massive and dwarf galaxies, likely depending on factors such as feedback strength, ISM properties, and timescale.
In massive galaxies, AGN feedback is generally negative on global scales, although spatially localized positive feedback can occur.
Numerical simulations have shown that AGN-driven outflows can enhance star formation in local regions or on short timescales, particularly in disk-like ISM where feedback pressurizes dense gas and promotes clump formation \citep{Gaibler2012, Dugan2017,Mukherjee2018, Mercedes2023}. 
In this work, we demonstrate that AGN feedback can instead be globally positive in starburst dwarf galaxies. The key physical reason is that lower black hole masses drive relatively mild outflows that compress the dense ISM without expelling it, while the high gas densities in the starburst environment ensure efficient radiative cooling that rapidly converts shocked gas into star-forming conditions. Although this behavior may not be universal to all dwarf galaxies, our results suggest that it is a natural outcome in gas-rich starburst systems with efficient cooling and moderate AGN feedback.

This positive feedback mechanism may have important implications for understanding the evolution of high-redshift galaxies. 
The galaxies at high redshifts are observed to be more compact and gas-rich than their local counterparts \citep{Bezanson2009, Tacconi2010, Wang2022, Greene2024}, which may create favorable conditions for positive AGN feedback to operate.
Recent observations have also revealed an excess of massive galaxies at high redshifts compared to theoretical predictions \citep{Steinhardt2016, Carnall2023, Harikane2024}, which challenges our current understanding of galaxy formation and evolution. 
In addition, a high fraction of AGN has been identified in these early galaxies \citep{Maiolino2024, Fujimoto2024}. These findings raise the possibility that positive AGN feedback may be one of the plausible mechanisms contributing to the early fast formation of massive galaxies at high redshift \citep{Silk2024}.
Support for this scenario has also emerged from numerical simulations. For example, \cite{Zana2022} showed in a cosmological zoom-in simulation that quasar outflows can enhance the SFR of nearby satellite galaxies. Their model also reproduces the observed abundance of satellite galaxies around the most far-infrared (FIR) luminous quasars more successfully than simulations without AGN feedback. These results suggest that AGN-driven outflows may play an important role in shaping star formation in dense high-redshift environments.

In this work, we aim to investigate the role of AGN feedback in enhancing star formation in dwarf galaxies using high-resolution numerical simulations.
Our simulations are carried out with Multiscale AGN-regulated Cosmic Ecosystem Resolver in 3D (MACER3D) \citep{Zhang2025}, which is a three-dimensional hydrodynamic framework upgraded from its two-dimensional predecessor -- Massive AGN Controlled Ellipticals Resolved (MACER) \citep{Yuan2018}. Unlike cosmological-scale simulations, our code focuses on the galactic scale, enabling high spatial resolution and direct resolution of the Bondi radius, which is the outer boundary of the central black hole's accretion flow. 
Therefore, the gas accretion rate at the Bondi radius can be accurately calculated, which is essential for modeling realistic AGN feedback. 
Combined with the black hole accretion theory, a robust estimation of the accretion rate at the BH horizon can be obtained, which determines the outputs of AGN feedback, including radiation, jet and wind. By adopting advanced AGN physics, AGN feedback is divided into two modes, the hot (radio) mode and the cold (quasar) mode, depending on the black hole accretion rate. For the cold mode, the properties of AGN winds are adopted from observations of luminous AGN \citep{Gofford2015}. For the hot mode, the properties of AGN winds are adopted from theoretical models of hot accretion flows \citep{Yuan2015}. The radiative efficiency of hot accretion flow, which is different from that of standard thin disk, is adopted from theoretical models \citep{Xie2012}. Jet is not included in this work for simplicity, which will be investigated in future work. Crucially, the interaction between AGN outputs and the ISM is calculated directly in our simulations, avoiding parameterized approximations typically adopted in cosmological simulations.
Furthermore, our simulated dwarf galaxies are starburst systems with high gas fractions, consistent with observations that a significant portion of isolated low-mass galaxies are gas-rich \citep{Bradford2015}. 
We specifically focus on starburst dwarf galaxies because they provide the ideal environment where positive AGN feedback can be most clearly observed and distinguished from negative feedback effects. The high gas densities and efficient cooling in starburst environments create conditions where AGN-driven compression can significantly enhance rather than suppress star formation. By examining this specific regime, we can better understand the conditions under which AGN feedback can become positive.

The structure of this paper is as follows. In \S~\ref{sec:methods}, we introduce the initial conditions and setup of the simulations. In \S~\ref{sec:results}, we present the results of the simulations. In \S~\ref{sec:discussions}, we provide a discussion comparing our simulations with observations and other numerical works and summarize our findings.

\section{Methods} \label{sec:methods}

The simulations in this work are conducted using MACER3D, a three-dimensional hydrodynamic framework built upon Athena++ that employs spherical coordinates ($r, \, \theta, \, \phi$). 
A comprehensive description of the code configuration, including the implementation of AGN feedback physics, star formation and stellar evolution, supernova feedback, radiative cooling and heating processes, and metal treatment, is provided in \cite{Zhang2025}, so we briefly summarize the key physics here and introduce the initial conditions for dwarf galaxies in this section.

\subsection{Initial conditions and simulation setup} \label{sec:IC}
Our simulation initial conditions comprise a stellar and gaseous disk, a stellar bulge, a circumgalactic medium (CGM), and a central black hole (BH) embedded at the center of the dark matter (DM) halo.
The background gravitational potential that affects the gas dynamics is:
\begin{equation}
\Phi=\phi_{\rm BH}+\phi_{\rm disk,*}+\phi_{\rm bulge}+\phi_{\rm DM}.
\end{equation}
The self-gravity of gas is neglected at present, which will be included in future work.
The entire system is initialized in hydrostatic equilibrium. Following \cite{Springel2005}, we adopt the following density profiles for the stellar and gas disks and the stellar bulge:
\begin{equation} \label{disk}
\rho_{\rm disk}(R,z)=\displaystyle\frac{M_{\rm disk}}{2\pi z_{0} R_{0}^2}\,{\rm exp} \left(-\displaystyle\frac{R}{R_{0}}\right)\,{\rm sech} ^2 \left(\displaystyle\frac{z}{z_{0}}\right),
\end{equation}
\begin{equation} \label{bulge}
\rho_{\rm bulge}(r)=\displaystyle\frac{M_{\rm bulge}}{2\pi}\,\displaystyle\frac{r_{0}}{r(r+r_{0})^3},
\end{equation}
where $R$ denotes cylindrical radius, $z$ is the height above the disk plane, $r$ is the spherical radius, $M_{\rm disk}$ and $M_{\rm bulge}$ are the total masses of the disk and bulge components, respectively, and $R_{0}$, $z_{0}$, and $r_{0}$ represent the corresponding scale lengths.
The gaseous disk is characterized by a scale length of $R_{\rm 0}=2.2 \, {\rm kpc}$ and a scale height of $z_{\rm 0}=0.2 \, {\rm kpc}$.
The stellar disk is more compact, with a scale length of $R_{\rm 0}=1.8 \, {\rm kpc}$ and an identical scale height of $z_{\rm 0}=0.2 \, {\rm kpc}$.
The stellar bulge has a scale length of $r_{\rm 0}=0.3 \, {\rm kpc}$.
The initial rotational velocity of the gas disk is set according to the gravitational force and pressure gradients so as to satisfy hydrostatic equilibrium.

\begin{figure}[t]
\centering
\includegraphics[width=0.8\columnwidth]{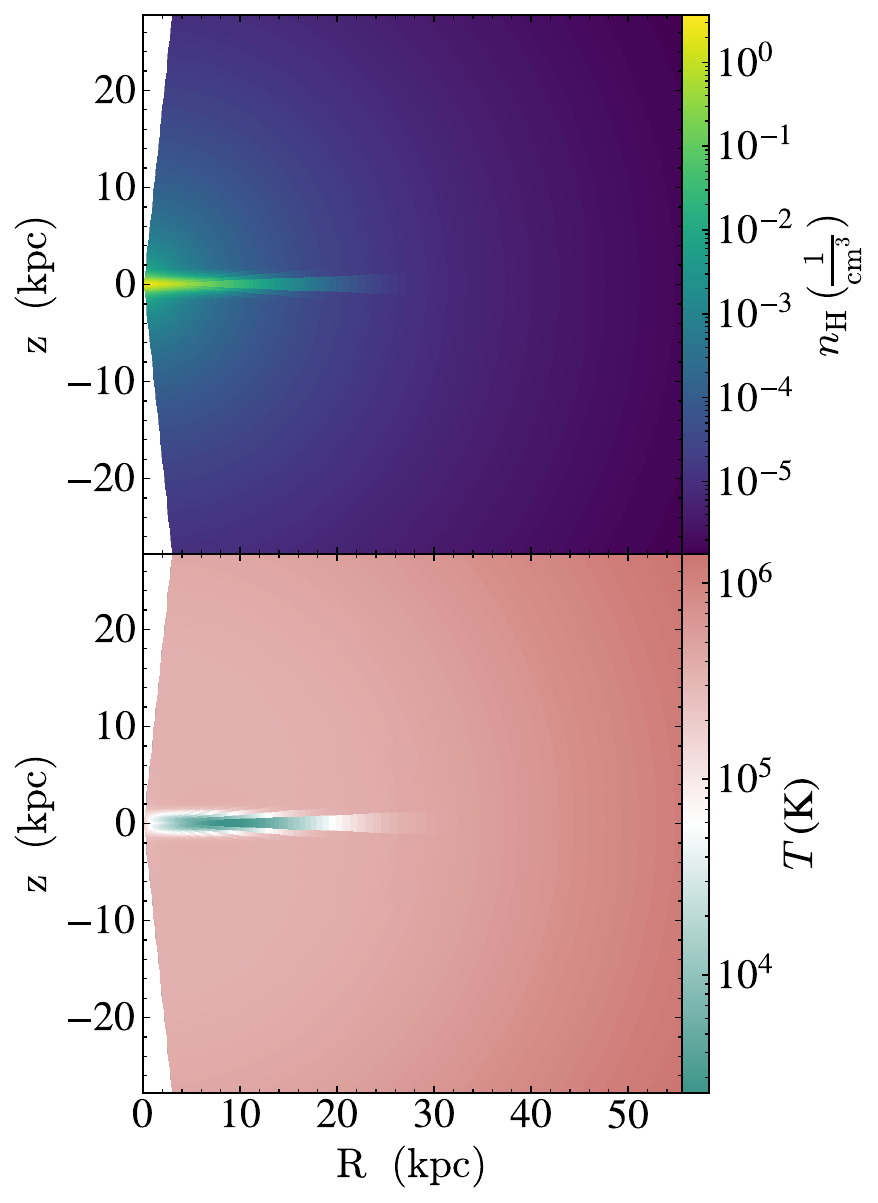}
\caption{Edge-on projections of the initial gas disk, showing gas number density (top) and temperature (bottom). The initial conditions satisfy hydrostatic equilibrium.
\label{IC}}
\end{figure}

We set the total stellar mass to $M_{*}=1.5 \times 10^{9}\, M_{\odot}$, distributed between the stellar disk ($M_{\rm disk, \, *}=1.25\times10^{9}\, M_{\odot}$) and the stellar bulge ($M_{\rm bulge}=2.5\times10^{8}\, M_{\odot}$). The gaseous disk has a mass of $M_{\rm disk, \, gas}=10^{9}\, M_{\odot}$. Based on the observed BH-stellar mass relation from \cite{Greene2020}, we assign a central BH mass of $M_{\rm BH}= 8.7 \times 10^{5}\, M_{\odot}$.
The CGM density follows a power-law profile, decreasing from approximately $10^{-4} \, {\rm cm}^{-3}$ in the inner regions to $10^{-6} \, {\rm cm}^{-3}$ at the outer boundary. The CGM temperature, derived from hydrostatic equilibrium considerations, is approximately $10^5 \, {\rm K}$.
The DM halo is represented by a Navarro-Frenk-White (NFW) profile \citep{Navarro1997} with a concentration parameter $c=10$, a total mass of $10^{10.8}\, M_{\odot}$, and a virial radius of $75 \, {\rm kpc}$.
The initial gas metallicity is set to $0.1 \, Z_{\odot}$ within the central region ($r < 0.125 \, R_{\rm 0, \, gas}$) and decreases to $0.01 \, Z_{\odot}$ at the outer boundary following a power-law distribution.
The initial conditions of the gas disk are shown in Fig.~\ref{IC}. The top panel presents the gas number density, while the bottom panel shows the temperature. The disk is initialized with smooth density and temperature profiles, without imposed density inhomogeneities.

Our computational domain spans from $7 \, {\rm pc}$ (corresponding to the Bondi radius of the central BH) to $75 \, {\rm kpc}$ (the virial radius of the galactic halo).
We employ a numerical resolution of $256 \times 64 \times 128$ grid cells, with logarithmic spacing in the radial direction that provides a finest resolution of $0.3 \, {\rm pc}$ at the inner boundary.
To avoid numerical singularities at the poles, the polar angle $\theta$ is restricted to the range $[8^{\circ}, 172^{\circ}]$.

We present five simulation configurations to investigate the interplay of different feedback mechanisms in dwarf galaxy evolution:
\begin{enumerate}
    \item {\tt\string fullFB}: includes both AGN and SN feedback;
    \item {\tt\string SNonly}: includes only SN feedback;
    \item {\tt\string AGNonly}: includes only AGN feedback;
    \item {\tt\string fullFB (inflow)}: incorporates both feedback mechanisms plus cosmological inflow;
    \item {\tt\string SNonly (inflow)}: includes SN feedback and cosmological inflow.
\end{enumerate}
We include cosmological gas inflow to represent the continuous external gas supply expected for dwarf galaxies embedded in the cosmic web. In the absence of such inflow, the galactic gas reservoir would be gradually depleted by star formation and feedback processes, leading to a decline in star formation activity over time.
Furthermore, we investigate how cosmological inflow influences galaxy evolution, specifically whether it modifies the impact of AGN feedback on star formation and regulates black hole accretion and AGN activity. These questions are discussed in this work.
The cosmological inflow is implemented via two filamentary structures entering from the outer boundary, each with a constant mass inflow rate of $1.5 \, {\rm M_{\odot} \, yr^{-1}}$ and an inflow velocity of $120 \, {\rm km \, s^{-1}}$, consistent with cosmological simulations \citep{Nelson2013, Keres2005}.

The mass evolution from different sources (ISM, cosmological inflows, AGN winds, stellar winds and ejecta of SNe Ia and SNe II) are tracked by passive scalars in our simulations.

\subsection{AGN feedback} \label{sec:AGN_feedback}
The AGN feedback in MACER3D is based on the subgrid model of \cite{Yuan2018}, operating in two modes separated by a critical accretion rate $\dot{M}_{\rm BH} \sim 0.02\,\dot{M}_{\rm Edd}$ \citep{Yuan2014}. Because the Bondi radius is resolved in our simulations, the Bondi accretion rate is computed directly from gas properties at the inner boundary without any boost factor.
In the cold (quasar) mode ($\dot{M}_{\rm BH} > 0.02\,\dot{M}_{\rm Edd}$), a standard thin disk forms with a radiative efficiency of $0.1$ and $L_{\rm bol} = 0.1\dot{M}_{\rm BH}c^2$; wind mass and velocity follow observationally calibrated scaling relations \citep{Gofford2015}. For super-Eddington accretion, the BH accretion rate and wind properties are taken from general relativistic radiative MHD simulations \citep{Yang2023}, yielding a characteristic wind velocity of ${\approx}\,0.15c$.
In the hot (radio) mode ($\dot{M}_{\rm BH} \leq 0.02\,\dot{M}_{\rm Edd}$), a hot accretion flow forms inward of a truncation radius; wind properties follow the theoretical prescriptions of \cite{Yuan2015}, and the radiative efficiency—which is much lower than that of a thin disk and depends strongly on the accretion rate—is adopted from \cite{Xie2012}. Full details are given in \cite{Zhang2025}.

\subsection{Star formation and supernova feedback} \label{sec:SF}
Star formation follows the Kennicutt-Schmidt (KS) law \citep{Kennicutt1998}:
\begin{equation} \label{sf}
\dot \rho_{\rm SF}=\displaystyle\frac{\eta_{\rm SF} \, \rho_{\rm gas}}{\tau_{\rm SF}},  
\end{equation}
where the star formation efficiency (SFE) $\eta_{\rm SF}=0.03$ and the star formation timescale
\begin{equation}
\tau_{\rm SF}={\rm max}(\tau_{\rm cool},\tau_{\rm dyn}), 
\end{equation}
with the dynamical timescale $\tau_{\rm dyn}={\rm min}(\tau_{\rm ff},\tau_{\rm rot})$, where
\begin{equation} 
\tau_{\rm ff}\equiv\sqrt{\displaystyle\frac{3\,\pi}{32\,G\,\rho}}, \qquad
\tau_{\rm rot}\equiv\sqrt{\displaystyle\frac{r\,\partial \Phi(r)}{\partial r}}.  \label{eq:tdyn}
\end{equation}
Gas is eligible for star formation when its density exceeds $1\,{\rm cm^{-3}}$ and temperature falls below $4 \times 10^4\,{\rm K}$. Once star formation occurs, the corresponding mass is removed from the gas component and added to the stellar component in the same grid cell.
A temperature floor of $10^4\,{\rm K}$ is imposed; molecular cloud formation is therefore not explicitly resolved, and star formation is estimated via the KS law as a subgrid proxy. The adopted SFE of $3\%$ is slightly higher than the typical value of $\sim 1\%$ for molecular clouds \citep{Krumholz2007, Evans2014} to account for the unresolved multiphase medium at the densities accessible in our simulations ($1$--$1000\,{\rm cm^{-3}}$).

Stellar evolution including stellar winds and SNe largely follows \cite{Ciotti2012}. SNe are modeled as discrete events: the number occurring per time step is drawn from a Poisson distribution with an expected rate set by the local SN rate (see \citealt{Zhang2025} for details). Each type Ia (type II) SN injects $1.4\,M_{\odot}$ ($16\,M_{\odot}$) of mass and $0.85\times10^{51}\,{\rm erg}$ of energy \citep{Sukhbold2016}. To mitigate the overcooling problem in dense regions, we adopt the momentum-based subgrid model of \cite{Martizzi2015}, distributing mass, momentum, and thermal energy within a coupling radius determined by local gas density and metallicity.

\subsection{Radiative cooling, heating and metallicity} \label{sec:radiation_metal}
Radiative cooling uses a five-dimensional table generated with Cloudy \citep{Ferland2017}, depending on gas temperature, density, metallicity, redshift, and AGN radiative flux; molecular cooling is not included. The effects of the UVB and CMB are captured through the redshift dependence. Cooling is integrated using the Townsend exact scheme \citep{Townsend2009}. Heating processes include photoionization and Compton heating from AGN radiation and the UVB. AGN radiative transfer is explicitly computed, while stellar radiation is not yet included \citep{Deng2024,Zhu2025}. Full details are given in \cite{Zhang2025}.

Total gas metallicity is tracked as a passive scalar advected with the gas flow. Metal enrichment from type Ia SNe, type II SNe, and stellar evolution is included, following the metal yields of \cite{Hopkins2018} and \cite{Nomoto2013}.

\section{Results} \label{sec:results}

\subsection{Enhanced star formation in AGN-hosting dwarf galaxies} \label{sec:enhanced_SF}

\begin{figure*}[t]
\centering
\includegraphics[width=0.85\textwidth]{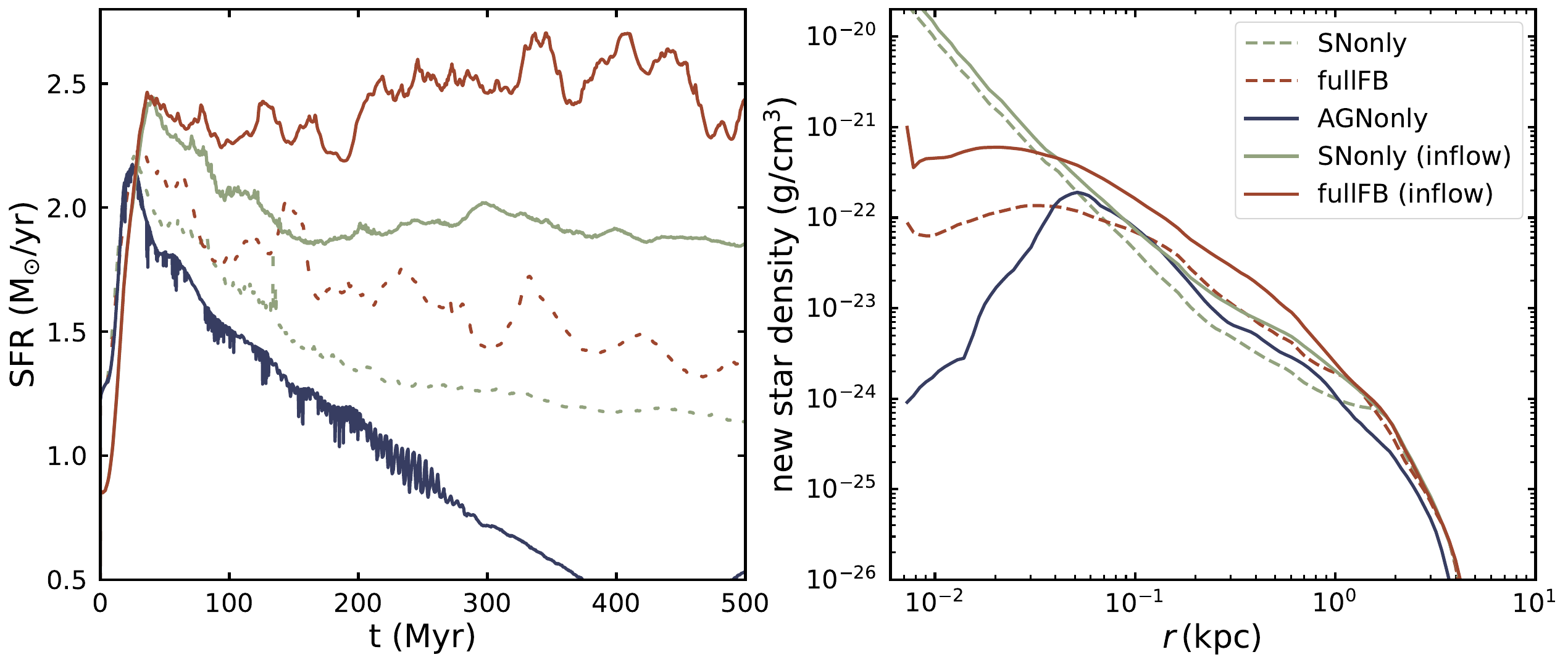}
\caption{Global SFR as a function of time (left) and profiles of the mass density of new stars at $t \, = \, 500 \, {\rm Myr}$ (right) in the {\tt\string fullFB (inflow)} (red solid), {\tt\string SNonly (inflow)} (green solid), {\tt\string fullFB} (red dashed), {\tt\string SNonly} (green dashed) and {\tt\string AGNonly} (grey) simulations. In {\tt\string fullFB (inflow)} and {\tt\string fullFB}, AGN feedback enhances global star formation: it suppresses star formation in the central region ($\lesssim 50\,\mathrm{pc}$) but promotes it across the wider galactic disk (from $\sim 50\,\mathrm{pc}$ to $1.5\,\mathrm{kpc}$). In the absence of SN feedback ({\tt\string AGNonly}), the star formation is instead significantly suppressed by AGN feedback.
\label{SFR_newstar}}
\end{figure*}

Our simulations reveal a striking and counterintuitive result: AGN feedback enhances rather than suppresses global star formation in starburst dwarf galaxies. Fig.~\ref{SFR_newstar} shows the evolution of SFR (left panel) and the radial profiles of the mass density of new stars (right panel) for different models. 
The mass density of new stars is calculated through the time integration of eq.~(\ref{sf}) over the entire simulation duration (500 Myr) at each radius.
As shown in the figure, models incorporating both AGN and SN feedback ({\tt\string fullFB} and {\tt\string fullFB (inflow)}) consistently maintain higher star formation rates compared to their counterparts with only SN feedback ({\tt\string SNonly} and {\tt\string SNonly (inflow)}). The {\tt \string fullFB (inflow)} simulation sustains an SFR of $\sim 2.4\,\mathrm{M_{\odot}\,yr^{-1}}$ -- approximately 25\% higher than the $\sim 1.9\,\mathrm{M_{\odot}\,yr^{-1}}$ in {\tt \string SNonly (inflow)}. These elevated SFRs (1.9-2.4 $\mathrm{M_{\odot}\,yr^{-1}}$) relative to the dwarf galaxy stellar mass ($1.5 \times 10^9 \mathrm{M_{\odot}}$) clearly place our simulated systems in the starburst regime.

The radial distribution of newly formed stellar mass (Fig.~\ref{SFR_newstar}, right panel) reveals that this enhancement is spatially structured: AGN feedback suppresses star formation in the central $\lesssim 50\,\mathrm{pc}$ region (due to localized heating and gas expulsion) but significantly enhances it throughout the galactic disk (from $\sim 50\,\mathrm{pc}$ to $1.5\,\mathrm{kpc}$). Importantly, this enhancement cannot be attributed simply to differences in gas reservoirs, as the total gas mass and cool gas mass in AGN feedback simulations exceed those without AGN by only 3\% -- 8\%, differences of which are too small to explain the observed 25\% increase in star formation.

This pattern challenges conventional wisdom about AGN feedback, which is typically associated with negative effects on star formation \citep{Croton2006,Silk2012,Dubois2013,Schawinski2007}. To understand this positive feedback mechanism, we must examine the physical conditions that enable it and the processes through which AGN activity enhances star formation in starburst dwarf galaxy environments. These mechanisms may be particularly relevant for understanding high-redshift, gas-rich dwarf galaxies during their most active formation phases.

\subsection{Shock compression of the ISM by AGN feedback with efficient cooling} \label{sec:shock_compression}

\begin{figure}[t]
\centering
\includegraphics[width=0.95\columnwidth]{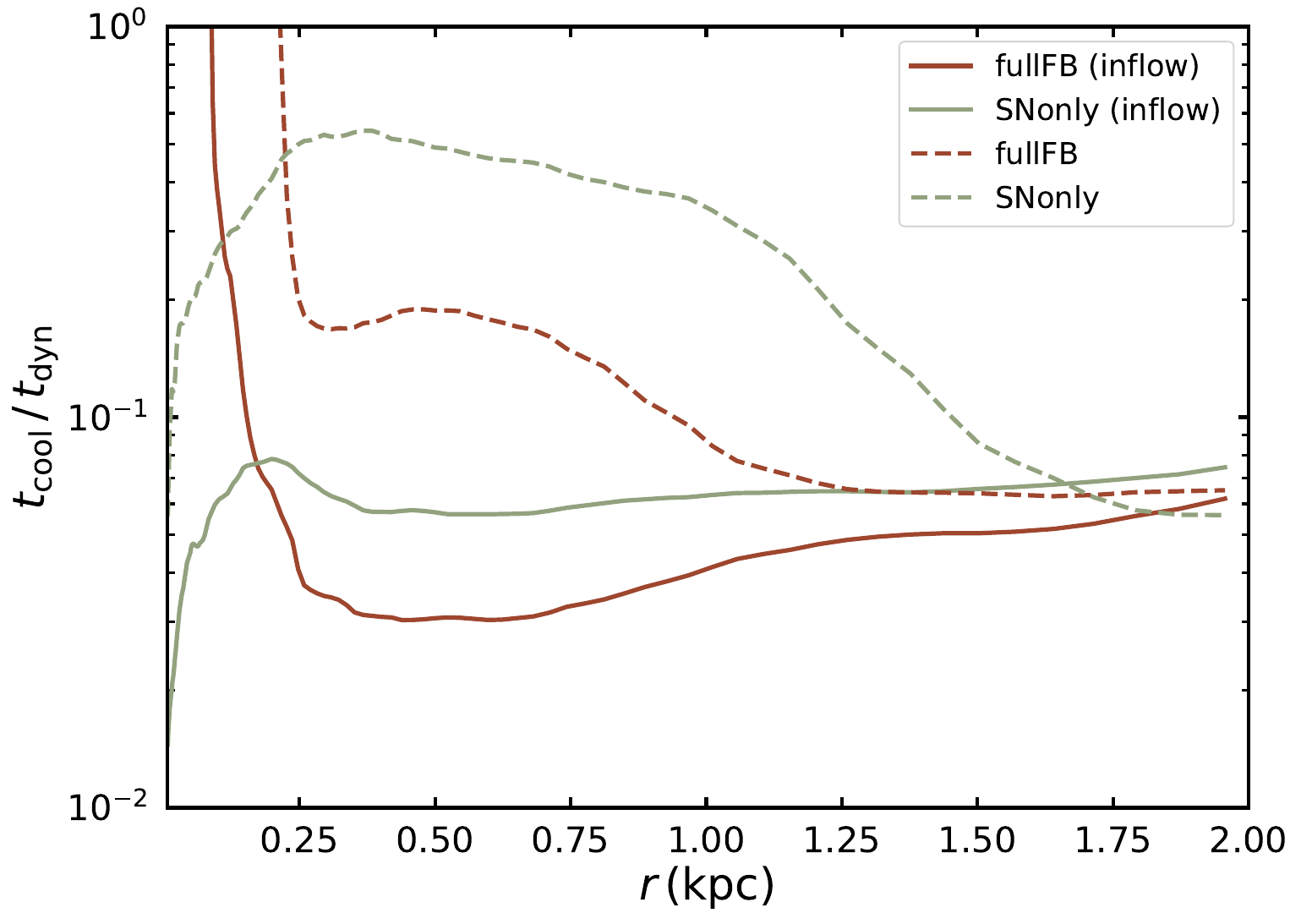}
\caption{Time-averaged radial profiles of the ratio of gas cooling timescale $t_{\rm cool}$ to its dynamical timescale $t_{\rm dyn}$. Within the region where star formation is most active ($\lesssim 2 \, \mathrm{kpc}$), the cooling timescale is shorter than the dynamical timescale, indicating efficient cooling in our simulated dwarfs.
\label{t_ratio}}
\end{figure}

To understand the physical reason of the enhancement of SFR, we have calculated the cooling and dynamical timescales, the results are shown by Fig.~\ref{t_ratio}, where the radial profiles are density-weighted azimuthal averages, and $t_{\rm dyn}$ is taken as the dynamical timescale defined in Eq.~\eqref{eq:tdyn}. This figure reveals a key feature of our simulated starburst dwarf galaxies: throughout the region where star formation predominantly occurs ($\lesssim$ 2 kpc), $t_{\rm cool}$ is consistently shorter than $t_{\rm dyn}$. This efficient cooling regime, characteristic of starburst environments, creates the essential conditions for positive AGN feedback. When gas is compressed by AGN-driven shocks, it initially heats up as kinetic energy is converted to thermal energy. The subsequent evolution depends critically on cooling efficiency, which has been explored in a series of numerical studies of the interaction between cold gas and hot winds \citep{Gronke2018, Kanjilal2021, Antipov2025}. In environments with efficient radiative cooling, as found in our simulated galaxies, this thermal energy is rapidly radiated away before the gas significantly re-expands. Consequently, the gas maintains its shock-compressed high density while quickly returning to lower temperatures that satisfy the star formation threshold. This creates a momentum-conserving radiative shock regime where the compressed gas remains dense rather than being dispersed by thermal pressure. In our simulations the post-shock gas cools efficiently, so the shocks are effectively radiative rather than adiabatic; the density jump therefore may exceed the factor-of-four adiabatic limit. The result is gas with both high density ($n \gtrsim 10\,\mathrm{cm}^{-3}$) and low temperatures ($T \lesssim 4 \times 10^4\,\mathrm{K}$), which are ideal conditions for accelerated star formation typical of starburst activity \citep{Silk2024}.

This efficient cooling regime stands in contrast to conditions typically found in more massive galaxies, where: (1) higher virial temperatures and lower gas densities result in longer cooling times; (2) more powerful AGN deliver greater energy per unit gas mass; and (3) deeper potential wells contain gas at higher initial temperatures. In such environments, energy-conserving flows predominate, leading to sustained heating of the ISM and potential gas expulsion -- the classical negative AGN feedback scenario. The efficient cooling in our starburst dwarf galaxies, combined with lower-energy input from smaller black holes, thus creates conditions where AGN feedback enhances rather than suppresses star formation. This mechanism may be particularly relevant for understanding positive AGN feedback in high-redshift gas-rich dwarf galaxies, where starburst conditions and efficient cooling are common.

\begin{figure*}[t]
\centering
\includegraphics[width=0.99\textwidth]{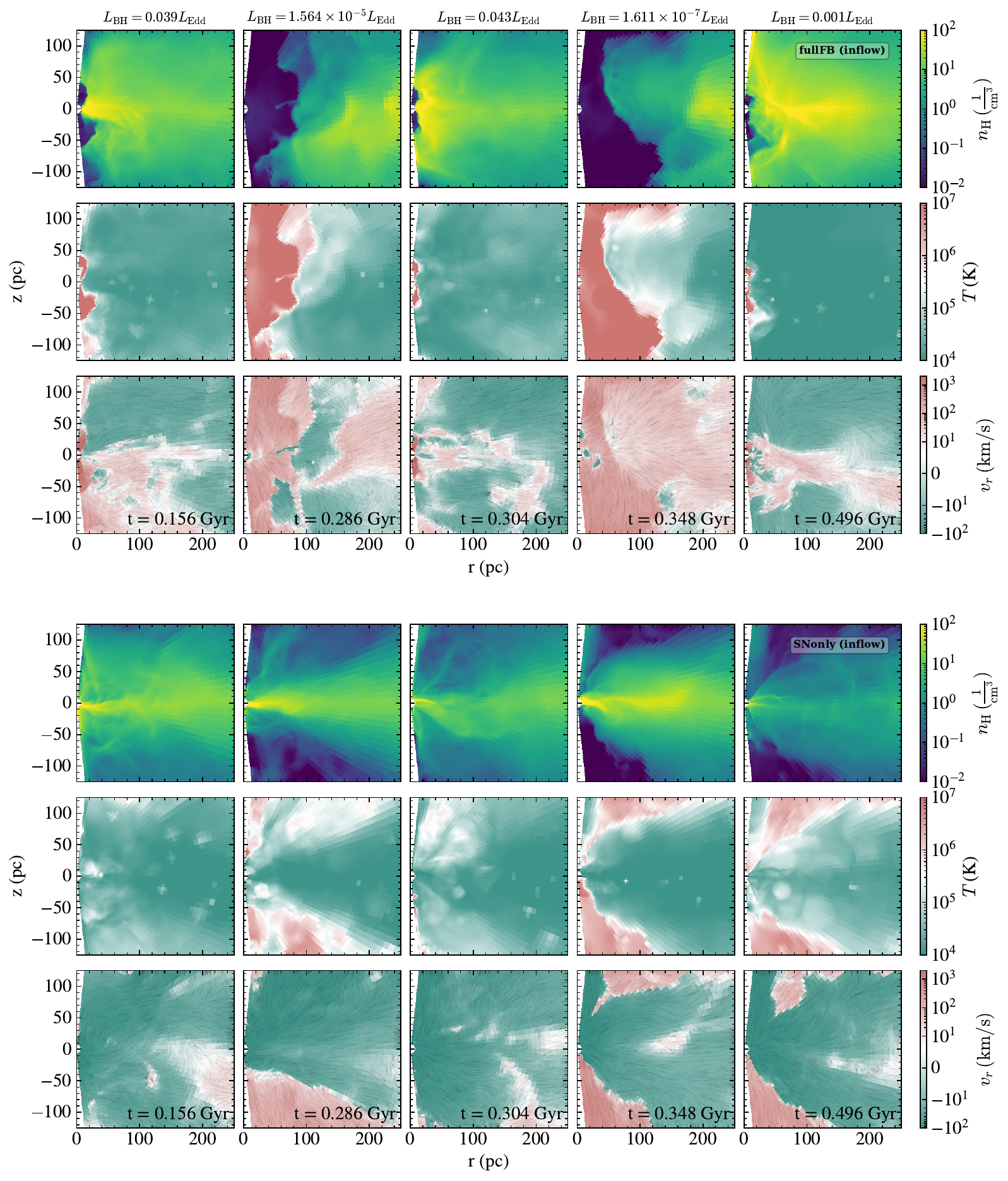}
\caption{Edge-on projections of various quantities within a 250 pc radius at different evolutionary stages in the {\tt\string fullFB (inflow)} (top panel) and {\tt\string SNonly (inflow)} (bottom panel) simulations: the first row shows volume-weighted gas number density; the second row displays mass-weighted gas temperature; the final row presents mass-weighted radial velocity with overplotted velocity streamlines. The AGN luminosities in the {\tt\string fullFB (inflow)} simulation and the corresponding simulation times are shown in each column. The compression of the ISM by AGN-driven shock waves produces dense gas structures, which in turn facilitate star formation.
\label{AGN_proj}}
\end{figure*}

\begin{figure*}[t]
\centering
\includegraphics[width=0.99\textwidth]{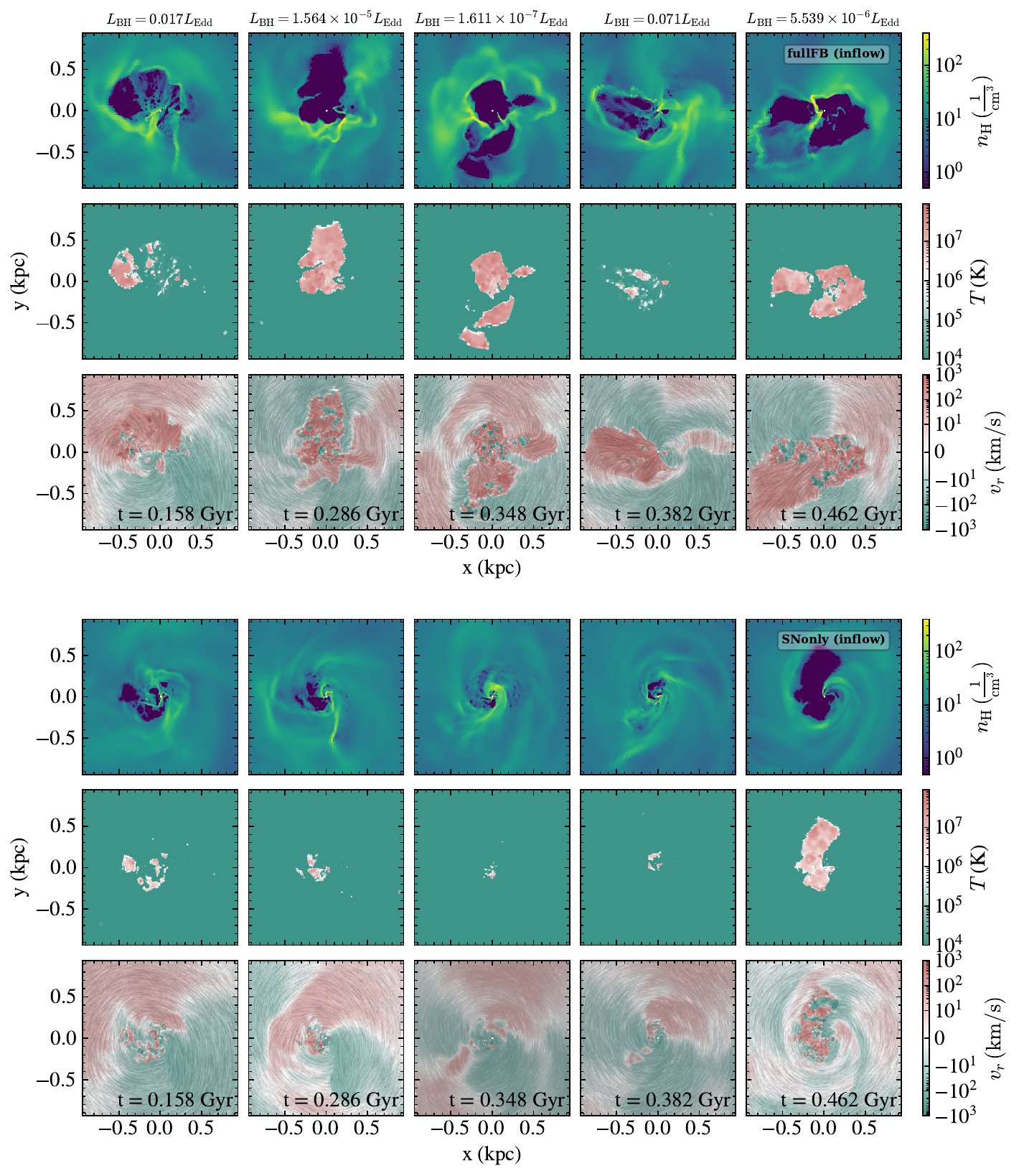}
\caption{Slice plots of gas density (first row), temperature (second row) and radial velocity (last row) on the equatorial plane within 1 kpc radius at selected evolutionary times for both the {\tt\string fullFB (inflow)} (top panel) and {\tt\string SNonly (inflow)} (bottom panel) simulations. Pronounced filamentary and clumpy structures are found in the {\tt\string fullFB (inflow)} simulation, which are largely absent in the {\tt\string SNonly (inflow)} case.
\label{slice}}
\end{figure*}

With efficient cooling established as the prerequisite condition, we now examine the mechanism through which AGN feedback enhances star formation. Fig.~\ref{AGN_proj} provides detailed projections comparing the {\tt\string fullFB (inflow)} simulation at different BH luminosities with the {\tt\string SNonly (inflow)} simulation at corresponding times. The volume-weighted gas density (top row), mass-weighted temperature (middle row), and mass-weighted radial velocity with streamlines (bottom row) collectively demonstrate the transformative impact of AGN feedback on gas structure.

During AGN active phases, strong outflows sweep up surrounding gas into compressed shells. Critically, AGN heating remains confined to the immediate vicinity of the black hole ($\lesssim 50$ pc), while the compressed shells extend much further. Within these shells, gas density increases significantly, while temperatures remain sufficiently low because of the short cooling timescale as shown in Fig. ~\ref{t_ratio}, which are ideal conditions for enhanced star formation. Even during quiescent phases of AGN, residual outflows continue to drive compression, maintaining elevated densities throughout the disk. In contrast, the simulation without AGN feedback shows no significant outflows within the gas disk and consequently lacks these compressed structures.

The broader impact of this compression mechanism is visible in Fig.~\ref{slice}, which presents equatorial plane slices within 1 kpc radius. The {\tt\string fullFB (inflow)} simulation exhibits pronounced filamentary and clumpy structures that are largely absent in the {\tt\string SNonly (inflow)} case. These dense structures form through two complementary processes: (1) direct compression of gas by AGN outflows; and (2) AGN outflows exploiting low-density channels created by supernova explosions to reach larger radii, where they collide with and compress the outer cool gas. This synergistic interaction between AGN and supernova feedback generates significantly more filamentary structures with substantially higher gas densities than observed in simulations with only supernova feedback. Quantitatively, the AGN in {\tt\string fullFB (inflow)} delivers energy at $\sim 10^{42}\,\mathrm{erg/s}$ during outbursts lasting $\sim 10\,\mathrm{Myr}$, with even a few percent coupling efficiency providing sufficient energy to compress gas within $\sim 0.5\,\mathrm{kpc}$ radius.

\subsection{Conversion of dense gas to enhanced star formation} \label{sec:dense_gas_SF}

\begin{figure*}[t]
\includegraphics[width=0.99\textwidth]{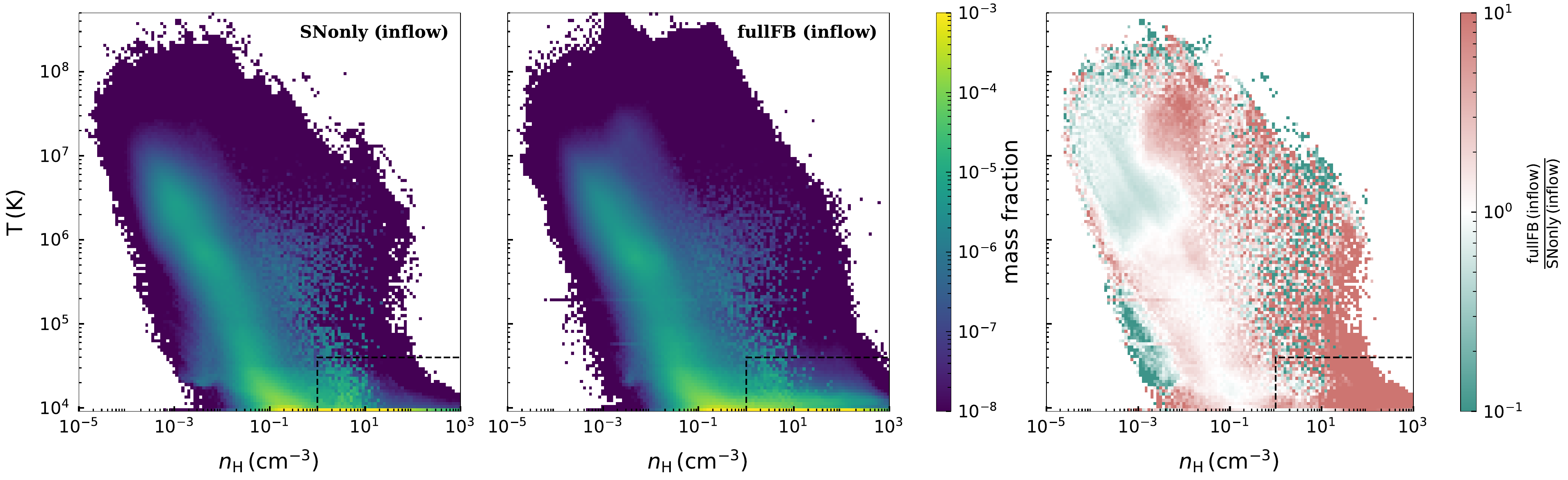}
\caption{\textit{Left and middle panels:} time-averaged temperature-density diagrams of gas within 2 kpc for the {\tt\string SNonly (inflow)} and {\tt\string fullFB (inflow)} models. \textit{Right panel:} the corresponding mass ratio between the {\tt\string fullFB (inflow)} and {\tt\string SNonly (inflow)} models. Unphysical infinite values have been removed. The black dashed rectangle represents the region containing star-forming gas. The {\tt\string fullFB (inflow)} simulation contains substantially more high-density gas than the {\tt\string SNonly (inflow)} simulations.
\label{phaseplot}}
\end{figure*}

The AGN-driven shock compression mechanism described above fundamentally alters the density distribution of gas in our simulated galaxies. Fig.~\ref{phaseplot} shows time-averaged temperature-density diagrams of gas within 2 kpc for the {\tt\string SNonly (inflow)} and {\tt\string fullFB (inflow)} simulations (left and middle panels) and the corresponding mass ratio of {\tt\string fullFB (inflow)} to {\tt\string SNonly (inflow)} models (right panel). The most striking difference appears in the high-density regime: the {\tt\string fullFB (inflow)} simulation contain substantially more gas with densities exceeding $10\,\mathrm{cm}^{-3}$ compared to the {\tt\string SNonly (inflow)} simulation. By the end of the simulation (t = 500 Myr), the {\tt\string fullFB} and {\tt\string fullFB (inflow)} simulations contain approximately three times more high-density gas than their counterparts with SN only.

This shift toward higher densities directly translates into enhanced star formation through the Kennicutt-Schmidt law implemented in our simulations:
\begin{equation}
\dot{\rho}_* = \epsilon_{\rm SF} \frac{\rho_{\rm gas}}{t_{\rm ff}} \propto \rho_{\rm gas}^{1.5}
\end{equation}
where $\dot{\rho}_*$ is the star formation rate density, $\epsilon_{\rm SF}$ is the star formation efficiency, $\rho_{\rm gas}$ is the gas density, and $t_{\rm ff} \propto \rho_{\rm gas}^{-0.5}$ is the free-fall time. This non-linear dependence on density ($\propto \rho_{\rm gas}^{1.5}$) means that even modest increases in gas density can produce substantial enhancements in star formation rates. We have explicitly verified this density-driven enhancement by decomposing the star formation contributions from different gas density regimes: we find that the enhancement in total SFR in simulations with AGN feedback is almost entirely attributable to increased star formation in high-density gas ($>10\,\mathrm{cm}^{-3}$). In contrast, star formation from intermediate-density gas ($1-10 \, \mathrm{cm}^{-3}$) shows negligible variation between the different feedback models. This analysis definitively links the enhanced star formation in our AGN feedback simulations to the increased presence of high-density gas created by AGN-driven shock compression.

\begin{figure*}[t]
\plotone{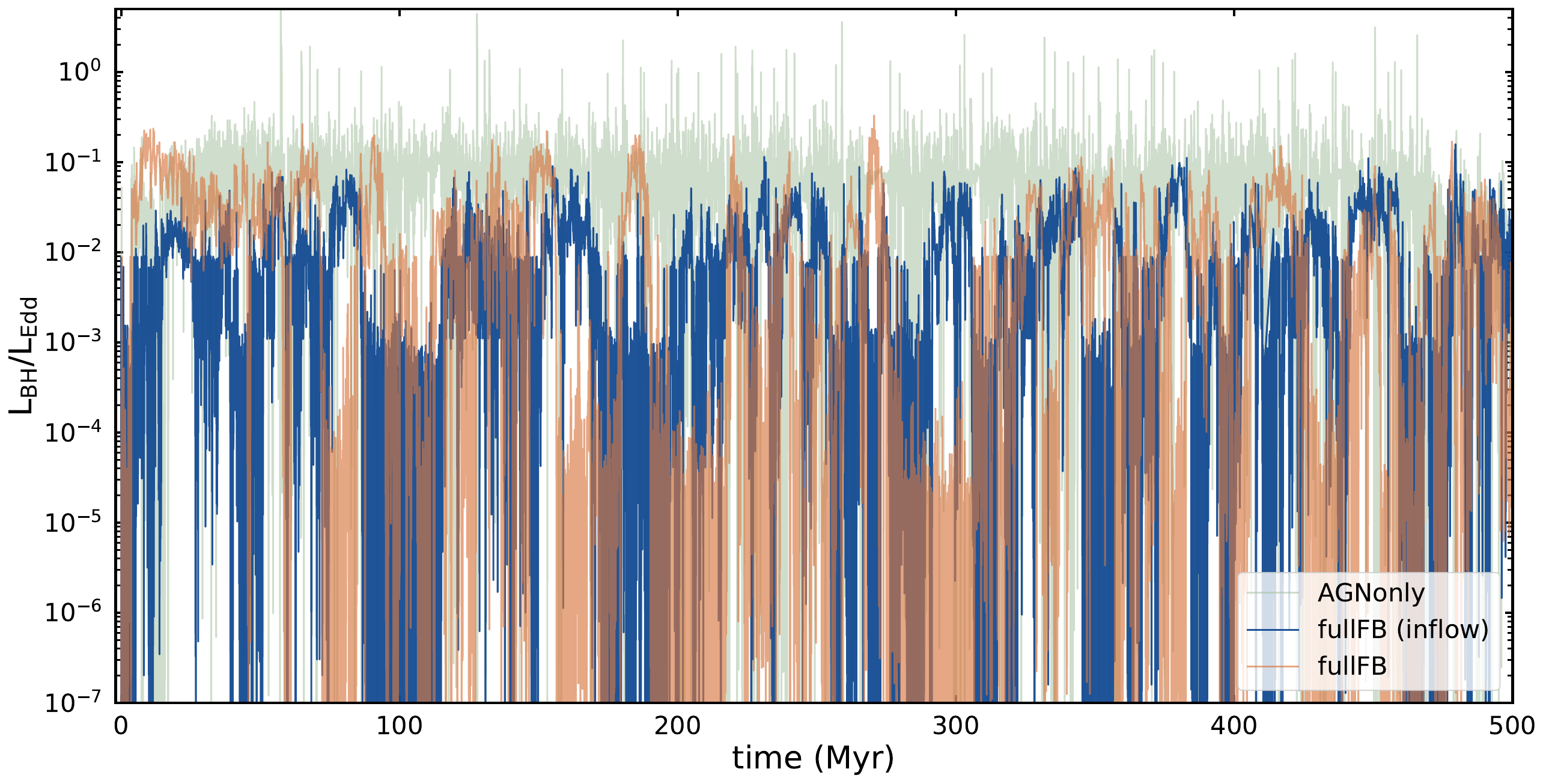}
\caption{Time evolution of BH luminosity in the {\tt\string AGNonly} (green), {\tt\string fullFB (inflow)} (blue) and {\tt\string fullFB} simulations (golden). The AGN luminosity in the {\tt\string AGNonly} simulation is significantly higher than that in the {\tt\string fullFB} simulation. The AGN luminosities in the {\tt fullFB (inflow)} and {\tt fullFB} simulations are comparable, indicating that cosmological inflow has little influence on central BH accretion.
\label{AGN_luminosity}}
\end{figure*}

Importantly, this mechanism operates effectively only when AGN feedback works in concert with supernova feedback. Additional simulations without supernova feedback ({\tt\string AGNonly})  revealed that AGN feedback alone does not enhance global star formation. The left panel of Fig.~\ref{SFR_newstar} shows the {\tt\string AGNonly} simulation exhibiting a much lower SFR compared to the other four. 
In the absence of SN feedback, higher accretion rates result in higher AGN luminosities (Fig.~\ref{AGN_luminosity}). SN feedback can reduce the gas density in the vicinity of BH, thereby suppressing accretion onto the BH.
In the {\tt\string AGNonly} simulation, the more powerful AGN expels gas from the disk rather than compressing it into star-forming structures. This underscores that the positive feedback mechanism requires the moderating influence of supernova feedback to function effectively in starburst dwarf galaxy environments. This synergistic feedback mechanism may have important implications for understanding the evolution of high-redshift gas-rich dwarf galaxies, which often exhibit both starburst activity and AGN signatures.

\subsection{Outflows across different galactic scales} \label{Galactic outflows}

\begin{figure*}[t]
\centering
\includegraphics[width=0.99\textwidth]{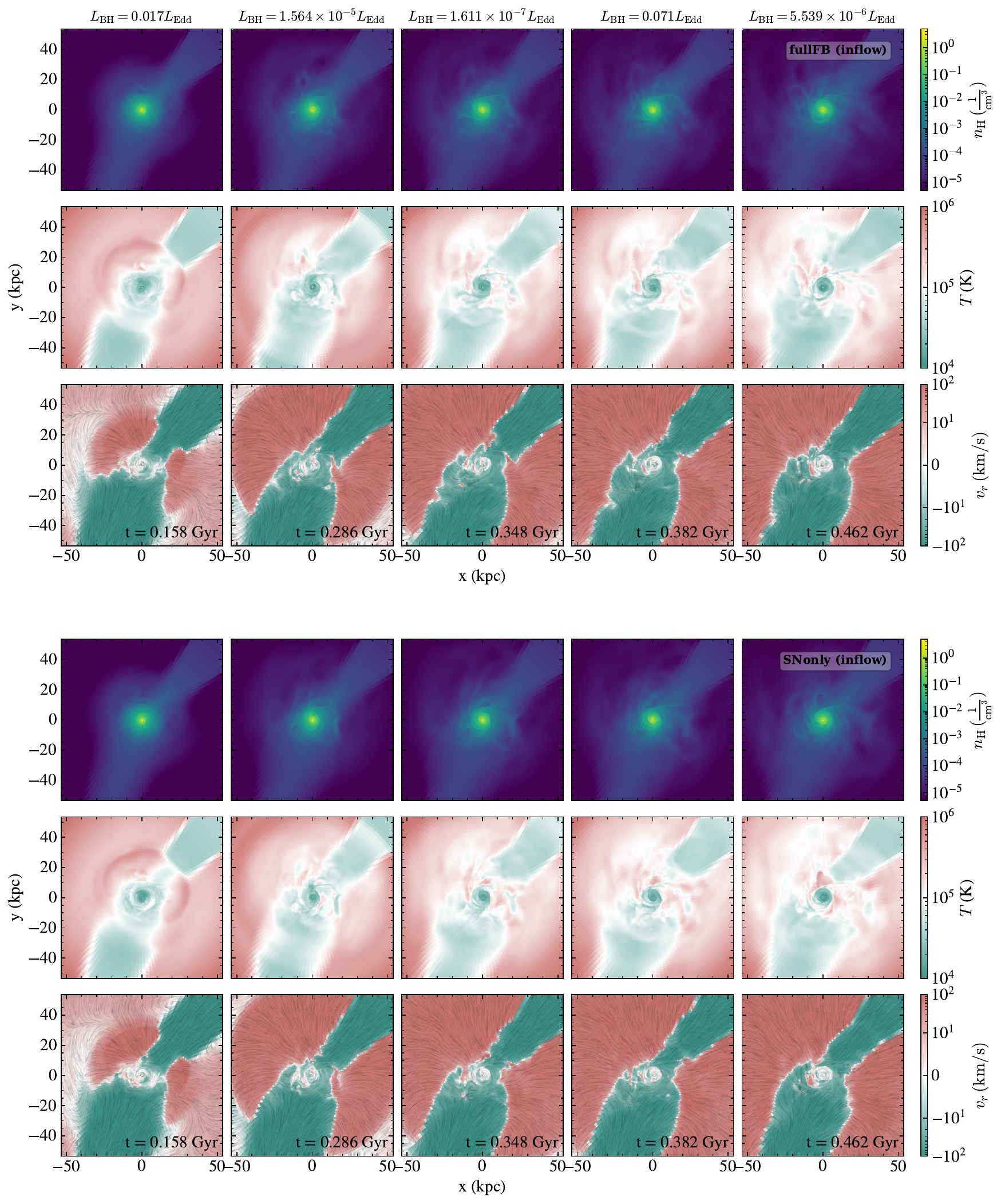}
\caption{Face-on projections of the same quantities as Fig.~\ref{AGN_proj}, zoomed out to a radius of 50 kpc at different evolutionary stages. The top panel corresponds to the {\tt\string fullFB (inflow)} simulation and the bottom panel corresponds to the {\tt\string SNonly (inflow)} simulation. The {\tt fullFB (inflow)} and {\tt SNonly (inflow)} simulations show no significant differences at CGM scale.
\label{faceon_inflow}}
\end{figure*}

Recently, some observations find that outflows from AGN-host low-mass galaxies are faster than those from low-mass star-forming/starburst galaxies without AGN signatures \citep{Liu2024, Salehirad2025}.
In our numerical simulations, switching AGN feedback on and off enables a clearer investigation of its impact on driving outflows in dwarf galaxies.

Fig.~\ref{faceon_inflow} presents the face-on projection plots zoomed out to a radius of 50 kpc, showing no significant difference between the {\tt\string fullFB (inflow)} and the {\tt\string SNonly (inflow)} simulations at CGM scale.
In the later stages of evolution, large-scale outflows with velocities reaching $ \sim \, 100 \, {\rm km/s}$ are observed in both simulations.
The mixing of galactic outflows with cosmological inflow and the CGM significantly reduces the CGM temperature, due to the interaction between galactic outflows, cosmological inflow, and the ambient CGM which naturally generates shear-driven turbulent mixing layers (TMLs; \citealt{Begelman1990,Ji2019,Tan2021,Yang2023b}).

\begin{figure*}[t]
\plotone{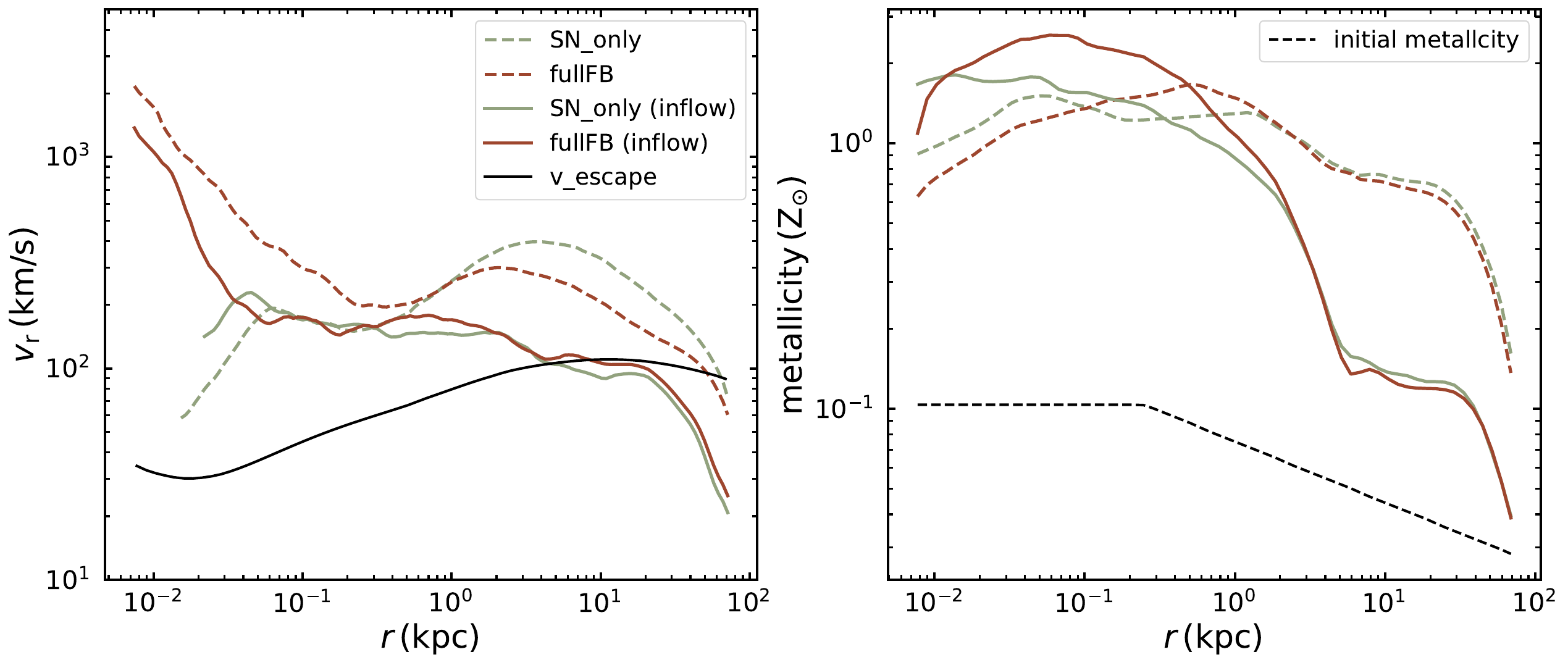}
\caption{The time-averaged radial profiles of outflow velocity ($v_{\rm r} > 0$) over the entire simulation period (left panel) and metallicity  during the last 200 Myr (right panel) for different models. The metallicity is calculated as the average over the last 200 Myr in order to assess the impact of outflows after a sufficiently long period. The statistical region is defined as $(\theta < 60^\circ)$ and $(\phi > 70^\circ)$, deliberately excluding the domain affected by cosmological inflow. In {\tt\string fullFB (inflow)} and {\tt\string fullFB}, AGN feedback can accelerate the outflow to velocities exceeding 1000 km/s in the galactic center ($r \lesssim 100\, {\rm pc}$). At larger radii, however, galactic outflows in these dwarf galaxies are dominated by SN feedback in all simulations. The mixing of cosmological inflow with galactic outflow reduces both the velocity and metallicity of galactic outflow.
\label{vr_metal}}
\end{figure*}

\begin{figure*}[t]
\includegraphics[width=0.99\textwidth]{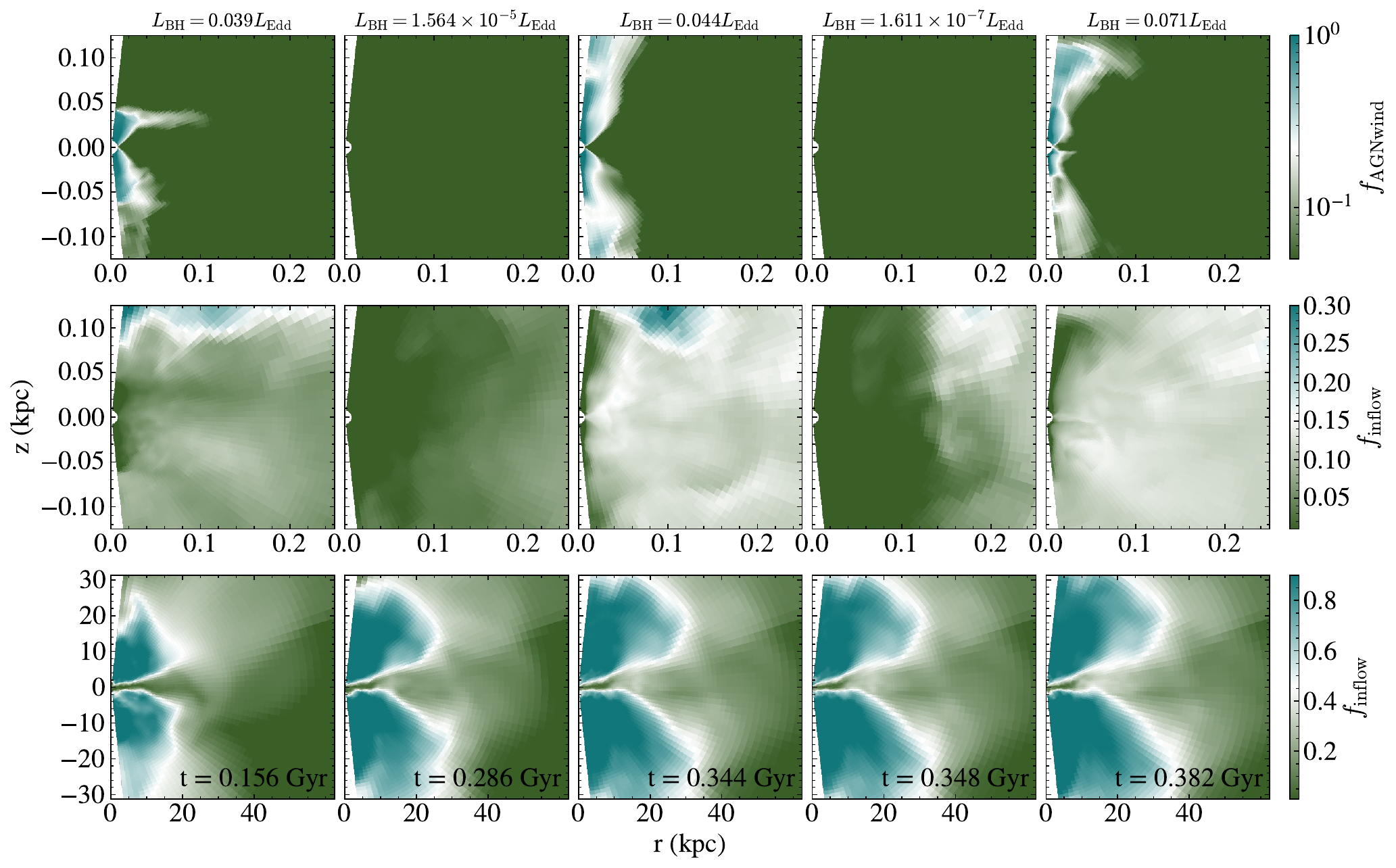}
\caption{Edge-on projections of AGN wind abundance (top row) and cosmological inflow abundance (middle and bottom rows) in the {\tt\string fullFB (inflow)} simulation at different simulated times. The top and middle rows display the central 250 pc region, while the bottom row shows the extended 60 kpc view. The AGN wind is confined in the galactic center ($r \lesssim 100\, {\rm pc}$). Inflowing gas is largely halted outside the galactic disk and contributes a small fraction to the central gas reservoir.
\label{fraction}}
\end{figure*}

The left panel of Fig.~\ref{vr_metal} shows the time-averaged radial profiles of outflow velocity over the entire simulation period in different models.
Only regions with radial velocity $v_{\rm r} > 0$ are included in the calculation.
In the central region ($r \lesssim 100\, {\rm pc}$), the outflow velocity (red lines: {\tt\string fullFB (inflow)} and {\tt\string fullFB}) is substantially higher than in the SN-only simulations (green lines: {\tt\string SNonly (inflow)} and {\tt\string SNonly}).
As shown by the AGN wind tracer (top row of Fig.~\ref{fraction}), AGN wind is also confined to this region, indicating that the central outflow in the {\tt\string fullFB (inflow)} and {\tt\string fullFB} simulations is primarily AGN-driven.
This small radius dominated by AGN feedback is consistent with previous order-of-magnitude estimates how far AGN feedback can effectively influence the ISM in dwarf galaxies in \S\ref{sec:shock_compression}.
AGN feedback can drive strong outflows in the central region, reaching velocities over 1000 km/s during active phases.
However, the velocities of AGN-driven outflow decreases rapidly with increasing radius. 
Beyond 1 kpc, the outflow velocities in the {\tt\string fullFB (inflow)} and {\tt\string fullFB} simulations converges to those in simulations without AGN feedback, as shown in the left panel of Fig.~\ref{vr_metal}.
The large-scale radial velocity projections in Fig.~\ref{faceon_inflow} show the same result.
These results indicate that galactic outflows in starburst dwarf galaxies are dominated by SN feedback.

As shown in the left panel of Fig.~\ref{vr_metal}, the SN-driven outflow velocity at larger radii in all simulations decline less rapidly than the AGN-driven outflows in the galactic center.
Because the AGN is a central point source, its outflows originate deep in the galactic potential well and encounter dense gas with efficient radiative cooling; consequently, AGN-driven outflows rapidly lose momentum and thermal energy and decay on small scales.
In contrast, SN-driven outflows are powered by supernovae occurring at spatially distributed sites throughout the galactic disk, rather than from a central source.
At larger radii, the shallower gravitational potential facilitates gas escape, and diffuse gas reduces radiative cooling losses. Therefore, the pressure gradients generated by hot, pressurized bubbles formed by supernovae at larger radii can more effectively drive gas outward.

We also find that the outflow velocities in simulations with cosmological inflow (solid curves) are lower than those without cosmological inflow (dashed curves), as shown in the left panel of Fig.~\ref{vr_metal}.
This velocity reduction does not result from suppression of AGN or SN feedback by the inflow.
Fig.~\ref{AGN_luminosity} shows comparable AGN luminosities in the {\tt fullFB (inflow)} and {\tt fullFB} simulations, indicating that cosmological inflow has little influence on central BH accretion.
This is further supported by Fig.~\ref{fraction}, where cosmological inflow is hindered outside the galactic disk (bottom row). 
Within a radius of 50 pc, the gas contributed by the inflow accounts for less than 15\% (middle row).
Consequently, the cosmological inflow fails to significantly fuel the central BH and have little effect on AGN feedback.
As shown in the left panel of Fig.~\ref{SFR_newstar}, the presence of cosmological inflow elevates the star formation rate, which in turn enhances SN feedback.
Therefore, as shown in Fig.~\ref{faceon_inflow}, the galactic outflow is suppressed as the cosmological inflow extensively mixes with outflows escaping from the galactic disk.

The colored curves in the right panel of Fig.~\ref{vr_metal} represent the time-averaged radial profiles of metallicity over the last 200 Myr for different models, excluding the inflow-injection region. The black dashed line indicates the initial metallicity profile.
For galaxies without cosmological inflows, galactic outflows transport more metals into the CGM, and can easily escape the galactic halo, further polluting the cosmic web.
When cosmological inflow is included, metal-poor inflow gas mixes with the metal-rich outflow, diluting the metallicity of outflow.
As a result, the CGM metallicity is less enhanced by galactic outflows, and the metallicity near the virial radius is close to the initial value.

\section{Discussions} \label{sec:discussions}

\subsection{Comparison with observations \label{vs. observations}}

Observational evidence of AGN feedback in dwarf galaxies has grown in recent years, providing crucial benchmarks for our simulation results. While AGN-driven outflows have been detected in multiple dwarf galaxies \citep{Zheng2023, Liu2024, Salehirad2025}, some observations suggest a negative feedback scenario where these outflows primarily heat and expel gas, thereby suppressing star formation \citep{Penny2018, Manzano-King2019}. 
\cite{Penny2018} reported that $\sim 10 \%$ of quenched dwarf galaxies in their sample host AGN, suggesting that AGN feedback may contribute to star formation suppression in dwarfs.
\cite{Manzano-King2019} observed that AGN-driven outflow velocities of dwarf galaxies in their sample are comparable to those in massive galaxies, and they suggested that such powerful outflows can expel gas from the shallow potential wells of dwarf galaxies, thereby inhibiting star formation.
Although the AGN-hosting dwarf galaxies in their sample are not quenched, they exhibit lower star formation rates than typical star-forming dwarf galaxies.
If AGN-driven outflows can suppress star formation, they should preferentially be observed in galaxies with low SFRs. However, AGNs are more readily identified in redder galaxies, and their detection in star-forming systems is hampered by contamination from stellar processes \citep{Manzano-King2019, Schutte2022}.
Therefore, AGN activity observed in dwarf galaxies with low SFRs does not necessarily constitute strong evidence for AGN-driven negative feedback.
Additionally, as discussed in \S\ref{sec:shock_compression}, the positive AGN-feedback effect in dwarf galaxies is more likely when radiative cooling of the ISM is efficient. Conversely, dwarf galaxies with low SFRs typically lack dense cold gas, thus radiative cooling is inefficient, favoring AGN-induced suppression of star formation.

Our simulation findings align closely with the existing observational evidence of positive AGN feedback in dwarf galaxies, particularly the well-studied case of Henize 2-10, which represents one of the most clear-cut examples of positive AGN feedback in the local universe. Significantly, Henize 2-10 is a starburst dwarf galaxy, matching precisely the type of environment we have modeled in our simulations and providing an excellent observational counterpart to validate our theoretical predictions.
Henize 2-10 represents an ideal observational counterpart to our simulated starburst dwarf galaxy across multiple key parameters. This starburst galaxy maintains an SFR of approximately $1.9 \, {\rm M_{\odot}/yr}$ \citep{Reines2011}, quite similar to our simulation results. The agreement extends to fundamental galactic properties, with Henize 2-10 hosting a black hole of mass $1.5 \times 10^{6} \, {\rm M_{\odot}}$ and stellar mass of $ 3.7 \times 10^{9} \, {\rm M_{\odot}} $ \citep{Riffel2020, Reines2011} -- both comparable to our simulated galaxy. These similarities in mass, star formation rate, and starburst classification make Henize 2-10 an exceptional real-world analog to validate our model predictions.

Given the intense star formation and highly sub-Eddington AGN in Henize 2-10, it is reasonable to question whether starburst-driven processes might contaminate the observations and thereby cast doubt on the inferred positive AGN feedback. For many years, there has been ongoing debate over whether the nuclear source is a massive black hole or a supernova remnant. However, advances in multiwavelength observations and observational facilities in recent years have provided compelling evidence for the presence of a low-accretion central black hole (see Table1 in \citealt{Schutte2022}). Additionally, Henize 2-10 is a nearby galaxy (9 Mpc) with a well-resolved central region and optical observations can achieve spatial resolutions of a few parsecs. An ionized filament and an elongated molecular gas structure are found to connect the region of the black hole with a site of recent star formation. The central ionized gas displays a sinusoidal feature in the position-velocity diagram reaching the star-forming region, consistent with a precessing bipolar outflow and unlikely to arise from supernova remnants. Elevated CO excitation is observed at the interface between the black hole outflow and the triggered star-forming region, indicating that the molecular gas is being compressed by the shock produced by the outflow from the BH. More detailed discussions of the observational evidence of positive AGN feedback in Henize 2-10 can be found in \cite{Schutte2022, Gim2024}.

Most significantly, the physical mechanism of positive feedback observed in Henize 2-10 closely matches our simulation results. \cite{Schutte2022} documented an ionized outflow from the central black hole with velocities of several hundred km/s that shocks the interstellar medium at approximately $70 \, {\rm pc}$ from the black hole. 
This shock compression creates conditions favorable for enhanced star formation, consistent with the mechanism identified in our simulations.
The luminosity of BH in Henize 2-10 is observed to be in a low state, with radio luminosity $L_{\rm R} \sim 4 \times 10^{35} \, {\rm erg \, s^{-1}}$ and X-ray luminosity $L_{\rm X} \sim 4 \times 10^{38} \, {\rm erg \, s^{-1}}$ \citep{Reines2012, Reines2016}, indicating highly sub-Eddington accretion (Eddington ratio $\sim 10^{-6}$).
As demonstrated in Fig.~\ref{AGN_proj} and Fig.~\ref{slice}, our model shows dense star-forming gas structures formed by AGN-driven outflow compression at similar distances during quiescent phases, with comparable outflow velocities of several hundred km/s.

This alignment between our simulation results and observations of Henize 2-10 provides strong validation for our model's prediction that AGN feedback can enhance rather than suppress star formation in starburst dwarf galaxies under specific conditions. Our work offers a theoretical framework that explains the positive feedback mechanism observed in Henize 2-10 and suggests this phenomenon may be more widespread in starburst dwarf galaxies and possibly in high-redshift gas-rich dwarf galaxies than currently recognized.

Our results suggest that positive AGN feedback at galactic scales is plausible in starburst dwarf galaxies (environments with dense, efficiently cooling ISM) and warrants further investigation in this class of systems. Since AGNs in dwarf galaxies are often heavily obscured during periods of intense star formation, direct observational evidence for AGN-triggered star formation in these systems remains rare. Owing to this limitation and the restricted spatial resolution of current observations, Henize 2-10 is currently the only well-established case in the local universe. Significantly, it is itself a starburst dwarf galaxy whose physical conditions closely match those modeled in our simulations, providing excellent observational support for our theoretical predictions. In addition, enhanced star formation created by AGN-driven outflows compressing the ISM has been identified in a number of massive galaxies, which share a similar physical mechanism of positive feedback \citep{Shin2019,Pak2023,Zhang2023,Venturi2023}.
With the advent of future high-angular-resolution facilities, it will become possible to resolve the multiphase structure of the ISM and directly trace the interaction between AGN-driven outflows and star-forming regions in more low-mass systems.

\subsection{Comparison with other numerical simulations \label{vs. simulations}}

Recent numerical studies have increasingly incorporated AGN feedback in dwarf galaxy simulations, with most reporting star formation suppression rather than enhancement \citep{Koudmani2019, Koudmani2022, Sharma2023, Arjona-Galvez2024, Bi2025}. Among these studies, only \cite{Hazenfratz2025} documented modest star formation enhancement due to AGN feedback in specific scenarios, while their other simulations exhibited negative feedback effects.

The methodological differences between our approach and previous studies may explain these contrasting findings. Most existing simulations employ the Bondi-Hoyle-Lyttleton accretion model \citep{Hoyle1939, Bondi1944, Bondi1952} due to insufficient resolution to resolve the Bondi radius directly. This simplified approach potentially compromises the accuracy of black hole accretion rate calculations.
Moreover, the AGN physics adopted in these works is often inaccurate; for example, they do not differentiate between the cold and hot modes or simply assume a fixed AGN wind speed.
In addition, these models typically parameterize AGN-ISM interactions rather than computing them explicitly from physical principles. 
A fixed thermal-energy coupling efficiency is often arbitrarily assumed to represent the complex interactions between AGN feedback and the ISM, such as radiative transfer and wind-ISM interactions.
The adopted coupling efficiency can vary by factors of several among different simulations, potentially leading to substantial differences in the resulting gas dynamics and star formation.

Our simulation framework offers several methodological advantages that may yield more realistic outcomes. By resolving the Bondi radius directly, we achieve more precise determination of black hole accretion rates. 
More accurate AGN physics \citep{Yuan2018} is also employed in our simulations.
Rather than parameterizing AGN-ISM interactions, we explicitly calculate photoionization and Compton heating from AGN radiation. Additionally, our model incorporates variable AGN-driven wind velocities based on observational data for cold mode outflows \citep{Gofford2015} and theoretical models for hot mode outflows \citep{Yuan2015}, with velocities that scale with black hole mass and accretion rate rather than remaining fixed.

Comparative analysis suggests that previous simulations generally overestimate AGN power under similar galactic conditions, potentially explaining their tendency to predict star formation suppression. For instance, the simulations by \cite{Hazenfratz2025} demonstrated positive feedback only with low-mass black holes ($\sim 10^3 \, M_{\odot}$) using fixed wind velocities of 3000 km/s, while their models with black hole masses comparable to ours ($\sim 10^6 \, M_{\odot}$) exhibited significant star formation suppression. These findings further suggest that conventional AGN feedback implementations may overestimate feedback strength in dwarf galaxy environments, highlighting the importance of physically realistic models that accurately capture the complex interplay between AGN activity and the interstellar medium in low-mass galaxies.

\subsection{Numerical convergence} \label{highres}

\begin{figure*}[t]
\includegraphics[width=0.99\textwidth]{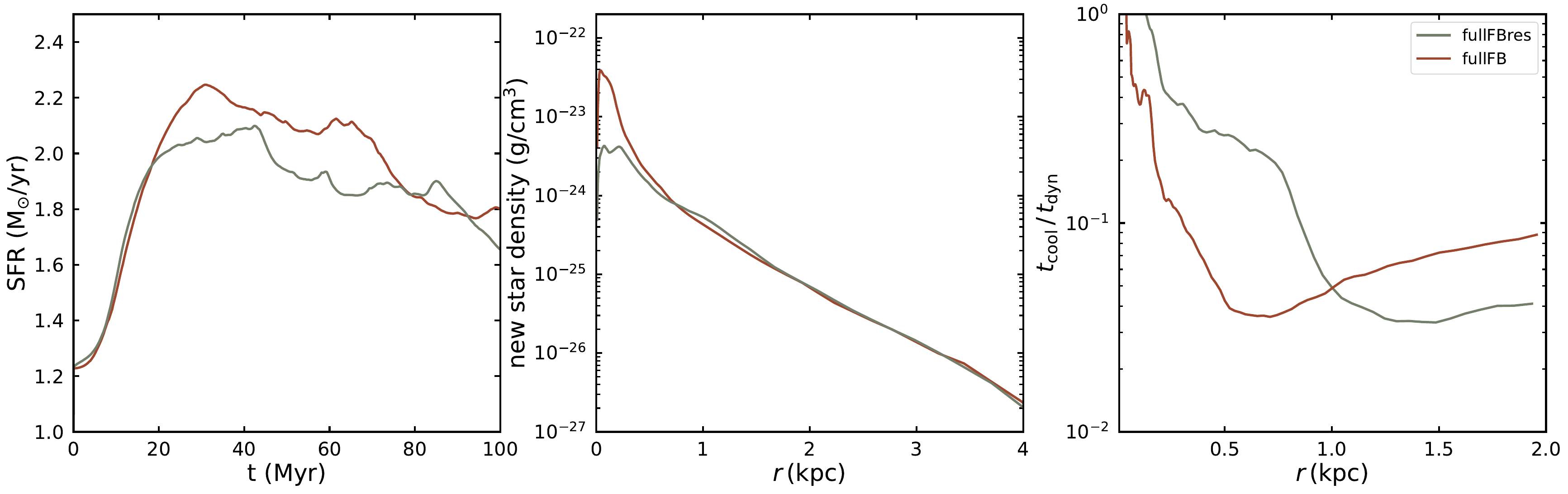}
\caption{Time evolution of SFR (left panel), radial profile of the newly formed stellar density at $t \, = \, 100 \, {\rm Myr}$ (middle panel) and the time-averaged radial profiles of the ratio of $t_{\rm cool}$ to $t_{\rm dyn}$ (right panel) over the period of $t \, = \, 60 - 100 \, {\rm Myr}$ in the {\tt\string fullFB} (red) and {\tt\string fullFBres} (green) simulations. The global SFR and spatial properties are broadly consistent, with $t_{\rm cool}/t_{\rm dyn}$ ratio remaining below unity in both runs which confirms that the efficient cooling regime is robust to resolution.
\label{SFR_newstar_tratio_highres}}
\end{figure*}

\begin{figure*}[t]
\plotone{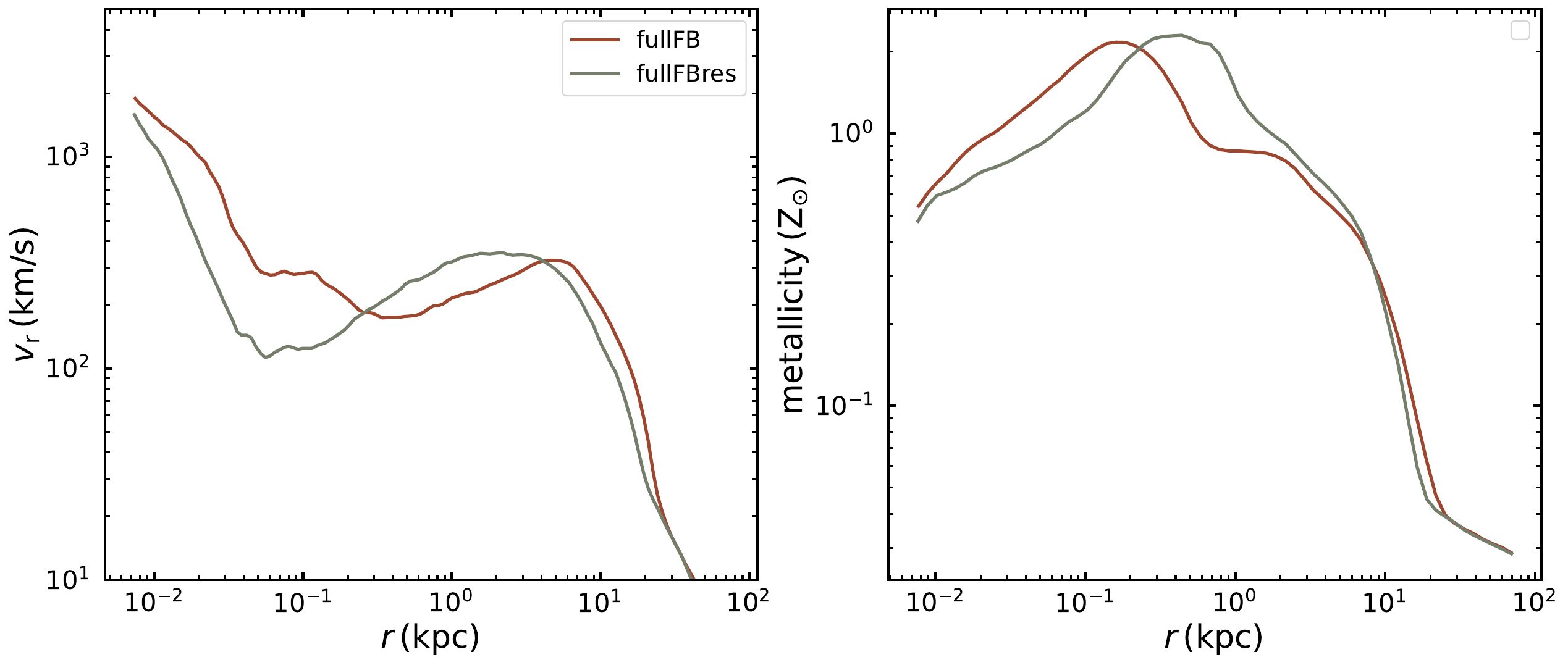}
\caption{The time-averaged radial profiles of outflow velocity (left panel) and metallicity (right panel) in the {\tt\string fullFB} (red) and {\tt\string fullFBres} (green) simulations over the period of $t \, = \, 60 - 100 \, {\rm Myr}$. The outflow velocity and metallicity distributions show good agreement between the two runs.
\label{vr_metal_highres}}
\end{figure*}

To assess numerical convergence, we performed a higher-resolution simulation {\tt\string fullFBres} with $512 \times 128 \times 256$ grid cells, reaching a spatial resolution of $\sim 0.2$ pc in the most central region. Due to the correspondingly small timestep (down to $\sim 10$ yr) imposed by the logarithmic radial grid, this run was evolved for 100 Myr.

As shown in Fig.~\ref{SFR_newstar_tratio_highres} and Fig.~\ref{vr_metal_highres} (cf.\ Fig.~\ref{SFR_newstar}, Fig.~\ref{t_ratio} and Fig.~\ref{vr_metal}), key properties such as the global SFR, the radial profiles of outflow velocity and metallicity are all broadly consistent between {\tt\string fullFB} and {\tt\string fullFBres}, confirming that our primary conclusions are robust to resolution. The newly formed stellar mass density profiles (middle panel of Fig.~\ref{SFR_newstar_tratio_highres}) show some reduction in the innermost region ($\lesssim 0.5$ kpc) at higher resolution, but this region occupies a very small volume on the spherical grid, and its contribution to the integrated global SFR is negligible, as reflected by the agreement in the left panel. The ratio $t_{\rm cool}/t_{\rm dyn}$ remains below unity at all relevant radii in both runs (right panel of Fig.~\ref{SFR_newstar_tratio_highres}), indicating that the efficient cooling regime that drives positive AGN feedback is not an artifact of resolution.

Although the overall trend is robust and the primary physical mechanism is captured, we note that the radial profile of $t_{\rm cool}/t_{\rm dyn}$ shows some quantitative differences within the inner $\sim 1$ kpc. The higher $t_{\rm cool}/t_{\rm dyn}$ in the inner region of the higher-resolution run may reflect resolution-dependent changes in the multiphase structure, including reduced numerical mixing between hot gas heated by feedback and the cooler ambient ISM. In addition, smaller-scale turbulent mixing layers \citep{Ji2019} and density fluctuations may be better captured at different resolutions, further changing the detailed spatial distribution of gas phases and thus the cooling properties. This highlights the complex interplay between resolution, multiphase structure, and cooling in these simulations, and suggests that while the overall positive feedback mechanism is robust, the detailed spatial distribution of gas phases and cooling properties may be sensitive to resolution effects.

\subsection{Conclusions \label{conclusions}}

In this paper, we investigated the role of AGN feedback in dwarf galaxies using high-resolution MACER3D simulations. Our results reveal that, contrary to conventional understanding, AGN feedback can significantly enhance rather than suppress star formation in starburst dwarf galaxies. This positive feedback mechanism, driven by shock compression with efficient cooling, provides new insights into the complex interplay between AGN activity and star formation in gas-rich dwarf environments. We find that AGN feedback enhances global star formation rates by approximately 25\% in starburst dwarf galaxies, operating through AGN-driven shock compression of the ISM combined with efficient radiative cooling, which creates dense gas structures conducive to star formation while quickly radiating away thermal energy.

The positive feedback exhibits clear spatial structure, suppressing star formation in the central region ($\lesssim 50$ pc) while enhancing it throughout the wider galactic disk (from $\sim 50$ pc to 1.5 kpc). Importantly, this mechanism requires both AGN and SN feedback working in concert; our additional tests revealed that AGN feedback alone would expel gas rather than compress it into star-forming structures. The moderate AGN energy input in dwarf galaxies compresses gas without expelling it from the shallow potential wells, creating ideal conditions for enhanced star formation when combined with efficient cooling.

Our simulation results may explain the observations of the starburst dwarf galaxy Henize 2-10, where AGN-driven outflows create shock-compressed regions of enhanced star formation at similar distances from the central black hole. This observational validation strengthens our conclusions about the positive feedback mechanism. While AGN drives powerful outflows in the central $\sim 100$ pc, we find that SN feedback dominates large-scale galactic outflows in dwarf galaxies, with outflow velocities exceeding 100 km/s, highlighting the complex interplay between different feedback mechanisms across galactic scales.

These findings have important implications for understanding both nearby starburst dwarf galaxies and high-redshift gas-rich dwarf galaxies. The efficient cooling regime characteristic of these environments creates ideal conditions for positive AGN feedback, suggesting that AGN may play a previously unrecognized role in accelerating star formation during the most active phases of dwarf galaxy evolution. This challenges conventional wisdom about AGN feedback and offers new perspectives on how black holes interact with their host galaxies in the low-mass regime.

\begin{acknowledgments}
We thank the anonymous referee for their constructive comments that improved the clarity of this paper. Authors are supported by the NSF of China (grants 12192223, 12522301, 12133008, and 12361161601), the China Manned Space Program (grants CMS-CSST-2025-A08 and CMS-CSST-2025-A10), the National Key R\&D Program of China No. 2023YFB3002502, and the National SKA Program of China (No. 2025SKA0130100). This work was performed in part at the Aspen Center for Physics, which is supported by National Science Foundation grant PHY-2210452. Numerical calculations were run on the CFFF platform of Fudan University, the supercomputing system in the Supercomputing Center of Wuhan University, and the High Performance Computing Resource in the Core Facility for Advanced Research Computing at Shanghai Astronomical Observatory.
\end{acknowledgments}

\software{{\small Athena++} \citep{Athenapp2020},
          {\small Cloudy} \citep{Ferland2017},
          {\small Matplotlib} \citep{hunter2007matplotlib},
          {\small NumPy} \citep{2020NumPy-Array}, 
          {\small SciPy} \citep{2020SciPy-NMeth}, 
          {\small yt} \citep{Turk2010,turk2024introducing}
          }

\bibliography{ref}{}
\bibliographystyle{aasjournalv7}

\end{document}